\documentclass[reprint,aps,prx,superscriptaddress,showpacs,nofootinbib,citeautoscrip,floatfix]{revtex4-2}

\usepackage{graphicx}
\usepackage{dcolumn}
\usepackage{bm}
\usepackage{flafter}
\usepackage{float}
\usepackage{lettrine}
\usepackage{amsmath}
\usepackage{amssymb}
\usepackage{color}
\usepackage{epsfig}
\usepackage{graphicx}
\usepackage[symbol]{footmisc}

\graphicspath{{eps+pdf/}}

\begin{document}

\date{\today}

\title{High-pressure structural and lattice-dynamics study of Yttria-Stabilized Zirconia}
\author{Shennan Hu}
\affiliation {Department of Materials Science and Engineering, Guangdong Technion-Israel Institute of Technology, Shantou 515063, China}
\author{Baihong Sun}
\affiliation{Department of Materials Science and Engineering, Guangdong Technion-Israel Institute of Technology, Shantou 515063, China}
\affiliation{Department of Materials Science and Engineering, Technion-Israel Institute of Technology, Haifa 3200003, Israel}
\author{Wenting Lu}
\affiliation{Department of Materials Science and Engineering, Guangdong Technion-Israel Institute of Technology, Shantou 515063, China}
\affiliation{Department of Materials Science and Engineering, Technion-Israel Institute of Technology, Haifa 3200003, Israel}
\author{Shiyu Feng}
\affiliation{Department of Materials Science and Engineering, Guangdong Technion-Israel Institute of Technology, Shantou 515063, China}
\affiliation{Department of Materials Science and Engineering, Technion-Israel Institute of Technology, Haifa 3200003, Israel}
\author{Bihan Wang}
\affiliation{Deutsches Elektronen-Synchrotron (DESY), D-22603 Hamburg, Germany}
\author{Hirokazu Kadobayashi}
\affiliation{Japan Synchrotron Radiation Research Institute, Sayo, Hyogo 679-5198, Japan}
\author{Yuzhu Wang}
\affiliation{Shanghai Synchrotron Radiation Facility, Shanghai Advanced Research Institute, Chinese Academy of Sciences, Shanghai 201204, China}
\author{Xingya Wang}
\affiliation{Shanghai Synchrotron Radiation Facility, Shanghai Advanced Research Institute, Chinese Academy of Sciences, Shanghai 201204, China}
\author{Lili Zhang}
\affiliation{Shanghai Synchrotron Radiation Facility, Shanghai Advanced Research Institute, Chinese Academy of Sciences, Shanghai 201204, China}
\author{Bora Kalkan}
\affiliation{Advanced Light Source, Lawrence Berkeley Laboratory, Berkeley, California 94720, USA}
\author{Azkar Saeed Ahmad}
\email{azkar.ahmad@gtiit.edu.cn}
\affiliation{Department of Materials Science and Engineering, Guangdong Technion-Israel Institute of Technology, Shantou 515063, China}
\affiliation{Guangdong Provincial Key Laboratory of Materials and Technologies for Energy Conversion, Guangdong Technion-Israel Institute of Technology, Shantou 515063, China}
\author{Elissaios Stavrou}
\email{elissaios.stavrou@gtiit.edu.cn}
\affiliation{Department of Materials Science and Engineering, Guangdong Technion-Israel Institute of Technology, Shantou 515063, China}
\affiliation{Department of Materials Science and Engineering, Technion-Israel Institute of Technology, Haifa 3200003, Israel}
\affiliation{Guangdong Provincial Key Laboratory of Materials and Technologies for Energy Conversion, Guangdong Technion-Israel Institute of Technology, Shantou 515063, China}

\begin{abstract}
The structural evolution of two selected compositions  of Yttria-Stabilized Zirconia (YSZ), with 3mol\% (3YSZ) and 8mol\% (8YSZ) of Y$_2$O$_3$, have  been investigated  under pressure using  in-situ synchrotron X-ray diffraction (XRD) and Raman spectroscopy in a diamond anvil cell up to 40 GPa (at room temperature). The close crystallographic relation between  the observed structures and the relatively large difference in the atomic numbers of Y/Zr and O,  imposes the simultaneous study using  both techniques, aiming to fully elucidate  the  structural evolution under pressure. The results, by combining both techniques, reveal that for both 3YSZ and 8YSZ, pressure promotes higher-symmetry structures. Under initial compression, the minority at ambient conditions monoclinic phase (m-phase) gradually transforms towards t-phase, a transition that is concluded for both 3YSZ/8YSZ at $\approx$ 10 GPa.  At higher pressures,   the solely remaining t-phase of 3YSZ transforms to the t'', that in turns transforms to the c-phase above 28 GPa. Likewise, for 8YSZ the coexistence of t- and t''-phases continue up to  31 GPa, where both transforms towards c-phase, that remains stable up to the highest pressure of this study. Upon pressure release, all observed transitions are fully reversible with negligible hysteresis, with the exception of the practical disappearance of the monoclinic phase at ambient conditions.  Our study underscores the significance of simultaneously performing  and analyzing the results of both XRD and Raman spectroscopy  studies in relevant crystallographic systems. Moreover, it provides a route towards a "structural purification" of YSZ through the elimination of the  m-phase aiming to improve material properties.  
\end{abstract}

\maketitle

\section{Introduction}
Yttria-Stabilized Zirconia (YSZ) is of great scientific interest due to its practical applications. It is the most widely employed material in thermal barrier coatings (TBCs) owing to its excellent shock resistance, low thermal conductivity, and relatively high coefficient of thermal expansion \cite{Chen2006}. Its  material properties and practical use in industrial aplications depend on the amount of dopant Yttria (Y$_2$O$_3$) in the YSZ. For example, 3YSZ (3mol\%Y$_2$O$_3$ doped ZrO$_2$) is a major part of cutting tools and dental implants, because of its superb hardness and corrosion resistance \cite{Yanagida1996}, while 8YSZ (8mol\%Y$_2$O$_3$ doped ZrO$_2$) is a preferred electrolyte for Solid Oxide Fuel Cells (SOFC) due to its high ionic conductivity \cite{mahato2015}. Moreover,  is also being used in oxygen sensors and oxygen pumps due to its superior high-temperature ionic conductivity, that is recognized to be the best  among YSZs   \cite{Maskell2000, Riegel2002, Pham1998, SakibKhan1998, Badwal1992}.

Structurally, pure ZrO$_2$ exist in differently polymorphous. At ambient conditions, pure ZrO$_2$ crystalizes into a monoclinic structure (Space group: $P21/c$, (14)) \cite{McCullough1959},referred thereafter as m-phase),  while  a tetragonal (Space group: $P42/nmc$, (137), with Zr and O at the 2a (3/4, 1/4, 3/4) and 4d (1/4, 1/4, z) Wyckoff  positions (WP), respectively, referred thereafter as t-phase ) and a cubic (Fluorite-type, $Fm\overline{3}m$, (225), with Zr and O at the 4a (0, 0, 0) and 8c (1/4, 1/4, 1/4) Wyckoff  positions (WP), respectively, referred thereafter as c-phase) structures are only stable at high temperature \cite{Leger1993,Ohtaka2005}. In other words, temperature induces the increase  in symmetry of ZrO$_2$, resulting to a change of its  physiochemical properties and improved material properties making it more suitable for industrial applications. The stabilization of these high-symmetry phases is of considerable scientific interest and is commonly achieved through doping with yttria (Y$_2$O$_3$), yielding yttria-stabilized zirconia (YSZ) \cite{Kumar2021}.

In the case of  YSZ, apart from the well-known m-phase  of  ZrO$_2$, the introduction of Yttria stabilizes the  t-phase and the  c-phase at room temperature (RT), depending on the doping level, although the quantitative  link between dopant amount and observed structure is still under debate. We note that the t-phase can be regarded as a distorted Fluorite c-phase with half the number of formula units (Z$_t$=2 vs Z$_c$=4): a) The lattice parameter $a$ deviates from the ideal $\sqrt{2}$ $a$ value for a cubic structure and b) the position of O atoms slightly deviates (typical values of $\approx$ z=0.465) from the ideal (for Fluorite phase) 1/4,1/4,1/4 values, see Fig. 1.  Many studies have reported that YSZ becomes fully tetragonal above 3mol\% Y$_2$O$_3$, and the full stabilization of the  cubic structure is not typically achieved until the Y$_2$O$_3$ content reaches values $>$ 12mol\% \cite{Yashima1995, Du1991,Lamas2000}. The stabilization of the t- and c-phases in YSZ upon  Yttria doping  is probably related with the   introduces oxygen vacancies and lattice distortions due to the charge mismatch between Zr\textsuperscript{4+} and Y\textsuperscript{3+}\cite{Yashima1996}. 

\begin{figure}[h]
    \centering
    \includegraphics[width=\linewidth] {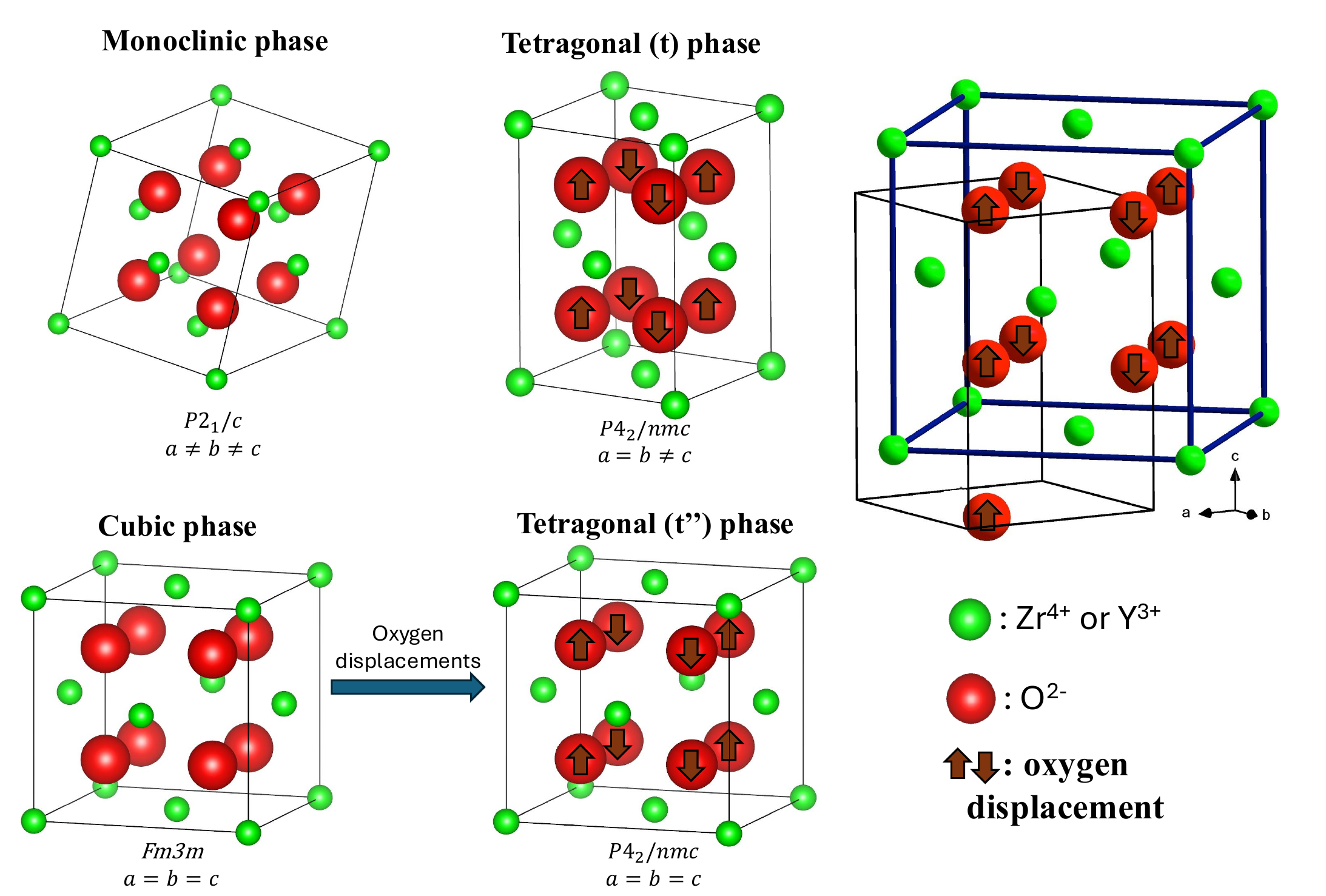}
    \caption{Schematic representations of the  monoclinic, tetragonal (t and t'', in the cubic-like representation) and the cubic crystal structures of YSZ adopted from  Yashima \textit{et al}. \cite{Yashima1996}. The arrows inside O atoms denote the displacement from the ideal (Fluorite-like) 1/4,1/4,1/4 positions. The relation between the tetragonal (black axes) and the distorted cubic fluorite-type (blue axes) crystal structures is also depicted. }
\end{figure}

In addition to above mentioned t- and c-phases, Yashima \textit{et al}. proposed the existence of additional tetragonal phases (t' and t'')(Space group: P42/nmc, (137)) \cite{Yashima1996, Yashima1993b}. The t'-phase maintains the tetragonal space group ($P4_2/nmc$) and  is referred to as  a metastable tetragonal phase to distinguish it from the stable t-phase \cite{Yashima1996}, which arises from cooling the high temperature c-phase and undergoes a diffusionless phase transition  at or below the equilibrium temperature \cite{gibson2001, Miller1981a, Scott1975, Yoshimura1990, zhou1991, Yashima1991}. Given that its crystallographic details are identical with  the t-phase, it will not be discussed in this study. The t''- phase retains the tetragonal symmetry, with respect to oxygen displacement,  but adopts the uniform lattice parameters of the c-phase, resulting in an axial ratio $c$/($\sqrt{2}$$a$) (tetragonality) equal to unity (1) \cite{Yashima1993}. In previous studies, the t''-phase has also been referred  as the c'-phase \cite{Yashima1993a, Yashima1993a}. 

These structural modifications critically influence the mechanical and ionic transport properties of YSZ. Understanding structural  behavior under extreme conditions is essential not only for optimizing YSZ-based industrial applications but also for advancing other doped oxide systems with similar oxygen-defect mechanisms (e.g. CeO\textsubscript{2}-Al\textsubscript{2}O\textsubscript{3}, ZrO\textsubscript{2}-Al\textsubscript{2}O\textsubscript{3}). Despite its importance, YSZ has been comparatively less explored under high-pressure. To date, only two brief high-pressure studies on YSZ have been reported. It should be stressed that certain experimental limitations exist in these studies, including non-hydrostatic pressure conditions or and relatively low pressure ranges \cite{Alzyab1987, smith2016},  which limit our understanding towards the behavior of the YSZ under pressure. 

Motivated by the above, in this study, we investigate the 3mol\% Y$_2$O$_3$-stabilized ZrO$_2$ (3YSZ) and  the 8mol\% Y$_2$O$_3$-stabilized ZrO$_2$ (8YSZ) up to 40 GPa. The 3YSZ composition lies at the stability limit of the tetragonal phase, where a coexistence with the monoclinic phase is observed, and 8YSZ  is known to adopt the t''-phase under ambient conditions \cite{Yashima1994}. We combined in-situ Raman spectroscopy together high-resolution synchrotron XRD mesurements using neon as the pressure-transmitting medium (PTM).

The results, by combining both techniques, reveal that for both 3YSZ and 8YSZ, pressure promotes higher-symmetry structures. At ambient conditions, in the case of 3YSZ, a coexistence of a, predominately, tetragonal  ($t$) and a monoclinic ($P2_1/c$) m-phase was observed. Upon initial pressure increase, the m-phase gradually transforms towards t-phase, a transition that is concluded  at $\approx$ 10 GPa. With further pressure increase, the tetragonal t-phase transforms to the t''-phase, followed by a transition to the c-phase above 28 GPa, that remains stable up to the highest pressure of this study. Likewise,  for 8YSZ a coexistence of the, predominately,  pseudo-cubic t''-phase and the m-phase was observed. Upon initial pressure increase, the m-phase gradually transforms towards t-phase, a transition that is concluded  at $\approx$ 10 GPa.  The coexistence of t- and t''-phases continue up to 31, where both transform towards the c-phase that remains stable up to the highest pressure of this study. Upon pressure release, all phase transitions are fully reversible with negligible hysteresis, with the exception of the practical disappearance  of the m-phase at ambient conditions. In this study, we emphasize that neither powder XRD nor Raman spectroscopy alone can uniquely resolve the pressure-induced phase transitions among t-, t''-, and c-phases in YSZ. Powder XRD is insensitive to oxygen displacements, while Raman spectroscopy cannot distinguish t- and t''-phases due to identical space group symmetry. 

\section{Experimental Methods}

\subsection{High-pressure study}
For all high-pressure studies, commercially available high-purity powder specimens of  3YSZ   (TCL,\>99.9\%) and  8YSZ (Adamas, \>99.9\%) with a nominal average grain size of 1 $\mu$m were used. High-pressure experiments are conducted using a mini-BX-80 Diamond Anvil Cell (DAC) with 300 or 400 $\mu$m in diameter culet diamonds. The sample chamber is prepared using  a rhenium gasket preindented to approximately 40 $\mu$m and  a central hole with diameter of 100$\sim$120 $\mu$m using  laser drilling. Pressure was determined by ruby fluorescence \cite{Syassen2008} and/or gold equation of state (EOS)  \cite{Anderson1989}. Neon was utilized as the PTM that filled the remaining space and was pre-compressed to $\approx$ 2KBar with a gas loading. Ne is known to remain fairly hydrostatic up to, at least, 40 GPa \cite{Klotz2009}.

\subsection{X-ray Diffraction}
At SPring-8, beamline BL10XU, a Flat Panel X-ray Detector (Varex Imaging, XRD1611 CP3)  was used, and the X-ray probing beam spot size was focused to about   10 x 10 $\mu{m}$  using compound refractive lens. More details on the SPring-8 XRD experimental setups are given in  Hirao  \textit{et al}. \cite{Hirao2020}.  At Beamline P02.2 at DESY, the X-ray probing beam was focused to a spot size of 2 x 2 $\mu{m}$ at the sample using Kirkpatrick-Baez mirrors and a PerkinElmer XRD 1621 flat-panel detector was used to collect the diffraction images of sample. A Dectris Pilatus3 S 1M Hybrid Photon Counting detector was used at the Advanced Light
Source, Lawrence Berkeley National Laboratory, Beamline 12.2.2. The spot size of the X-ray probing beam was focused to about 10 x10 $\mu{m}$ using Kirkpatrick-Baez mirrors. More details on the XRD experimental setups are
given in Kunz \textit{et al.} \cite{Kunz2005}. High-pressure XRD measurements were also performed at the Shanghai Synchrotron Radiation Facilities (15U and 17U beamlines).  

Integration of powder diffraction patterns to obtain intensity versus 2$\theta$ diagrams and initial analyses were conducted using the DIOPTAS software package \cite{Prescher2015}. Simulated XRD patterns were generated using the POWDER CELL program \cite{Kraus1996}, based on the experimentally determined EOSs and assuming continuous Debye rings with uniform intensity. Indexing of XRD patterns was performed using the DICVOL algorithm \cite{Boultif2004}, as implemented in the FullProf Suite. Rietveld refinements were carried out using the GSAS-II software \cite{Toby2013}.

\subsection{Raman spectroscopy}
Raman spectroscopy (RS) measurements were carried out using a custom-built confocal micro-Raman system with a 532 nm solid-state laser for excitation in backscattering geometry. The laser spot size was approximately 4 $\mu$m. Spectra were recorded with a spectral resolution of 2cm$^-$$^1$ using a single-stage grating spectrograph equipped with a CCD array detector. Ultra-low-frequency solid-state notch filters enabled spectral acquisition down to 10cm$^-$$^1$ \cite{Hinton2019}.

\section{Results}
To properly contextualize our findings, we first review the previous experimental investigations on this material system. As mentioned in the introduction, doping of ZrO$_2$ with Y$_2$O$_3$ stabilizes high-symmetry structures (tetragonal and cubic) as opposed to the monoclinic structure for pure ZrO$_2$. Among the limited high-pressure studies on YSZ, Alzyab \textit{et al}. investigated single-crystal specimens of 3YSZ ,  consisting of a mixture of $\approx$ 80\% monoclinic and 20\% tetragonal phases, with 4:1 methanol-ethanol as PTM.  Raman spectroscopy was used up to 6 GPa and reported a slow, gradual, and irreversible monoclinic-to-tetragonal phase transition  occurring above 4.2 GPa \cite{Alzyab1987}. In another study, Smith \textit{et al}. conducted non-hydrostatic ($i.e.$ no PTM) high pressure experiments on both 3YSZ and 8YSZ using powder XRD ranging from 0 to 26 GPa. Their results suggested that the tetragonal space group P4$_2$/\textit{nmc} provides the best structural fit for both 3YSZ and 8YSZ, from ambient conditions up to 26 GPa, indicating no phase transitions within this pressure range \cite{smith2016}. 

Moreover, we note the inherent limitations of  XRD techniques (especially in the case of powder samples)  in detecting slight oxygen displacements due to the significant X-ray scattering  cross-sections difference between oxygen (Z=8) and heavier cations (Zr, Z=40; Y, Z=39). Thus, although XRD measurements can provide info about the presence of t- or t''/c-phases, they cannot distinguish between the t''- and c-phases.   As noted by Lamas \textit{et al}., the t-phase exhibit a splitting of the (220)$_t$ and (004)$_t$  Bragg peaks \footnote[1]{In (400)$_f$, (220)$_t$, (004)$_t$: t indicates that these reflections are referred to the tetragonal cell(t); f indicates the reflections referred to the cubic cell.}, while the c- and t''-phases both exhibit a single merged peak—either (400)$_f$ from the cubic structure or the combined (220)$_{t''}$ and (004)$_{t''}$ peaks for the t''-phase\cite{Yashima1994}. This similarity arises from the second-order nature of the tetragonal-to-cubic phase transition and the group-subgroup relation between the two structures.  Thus, the exact determination of the critical pressures and high-pressure phases (pseudo-cubic t’’ or cubic) only from XRD results is ambiguous. 

Likewise, Raman spectroscopy measurements can provide information about the presence of the c- or t/t''-phases, however, they cannot distinguish between the t- and t''-phases, due to their identical crystallographic properties ($i.e.$ SG and WP). Thus, only the combination of both techniques (XRD for long range order and Raman spectroscopy for short range order) can accurately provide info about the critical pressures and high-pressure phases. This is the approach we follow in this study.

\subsection{XRD and RS measurements at ambient conditions.}

The  XRD patterns for 3YSZ and 8YSZ at ambient conditions are shown in Fig.2(a) and (b), respectively, in comparison with the calculated patterns based on the previously reported structures for YSZ \cite{Lamas2000,Tyrsted2014}; see details in the discussion above. For 3YSZ, the comparison documents the  t- phase as the dominant phase. However, in the case of  8YSZ, it is not clear if the structure is the t''- or the c-phase, from XRD results. For this reason, Raman spectroscopy measurements were performed and the results clearly clarify identify the presence of the t''- phase (together with minor amount of the m-phase), see Fig. 2(c). Overall, our ambient-condition results for both 3YSZ and 8YSZ are in agreement with numerous previous studies, demonstrating that an increase in Y$_2$O$_3$ doping level leads to higher-symmetry structures  \cite{MolinaReyes2018, Yashima1995, Du1991,Lamas2000, Yashima1996, Yashima1993b}.  

\begin{figure}[h]
    \centering
    \includegraphics[width=\linewidth] {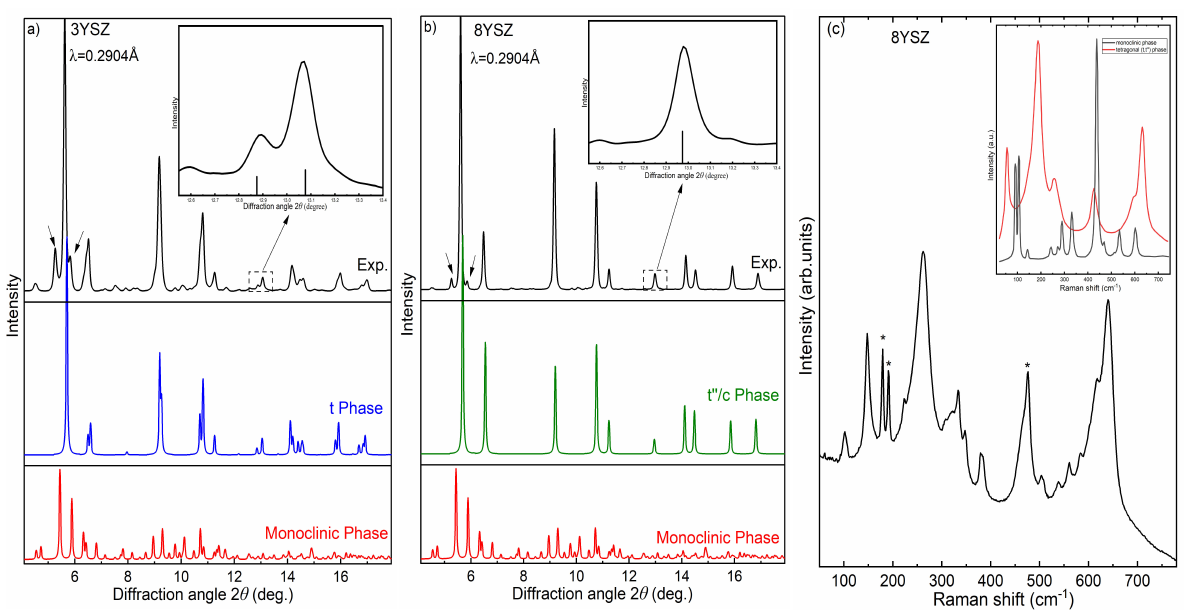}
    \caption{XRD patterns of (a) 3YSZ  and (b) 8YSZ at ambient conditions compared with relevant previously reported structures for YSZ\cite{Lamas2000, Tyrsted2014}; see text for details.  The downright arrows indicate the strongest Bragg peaks of the minority m-phase. The insets show an expanded view of the relevant patterns around 13$^{\circ}$ 2$\theta$ diffraction angle. (c) Raman spectrum of 8YSZ at ambient conditions, the asterisks denote most intense peaks of the m-phase. The inset shows the expected Raman spectra of the m- and t- phases of YSZ according to Ref. \cite{Hemberger2015}.}
\end{figure}

In both cases, the coexistence of the m-phase, as a minority phase, is observed.  In  Alzyab \textit{et al.} \cite{Alzyab1987}, a mixture of $\approx$80\% monoclinic and 20\% tetragonal phases was reported for  3YSZ at ambient conditions. As mentioned before, the exact quantitative  link between dopant amount and observed structure is still under debate, and probably synthesis conditions shall also play a role in the relative amount of observed phases. We also note that 8YSZ exhibits lower content of the m-phase due to the higher amount of doped Yttria.

\subsection{XRD and RS measurements under high pressure.}

\subsubsection{XRD measurements of 3YSZ up to 30 GPa}
Selected XRD patterns of 3YSZ at various pressures up to 30 GPa are shown in Fig. 3. Certain conclusions can be drawn from the XRD measurements of this study. First, upon increasing pressure, the m-phase gradually transforms into the t-phase, practically disappearing ($i.e.$ full transformation) above 10 GPa.  Alzyab \textit{et al.} reported a phase abundance of about 85\% tetragonal and 15\% monoclinic at 4.2 GPa for 3YSZ \cite{Alzyab1987}. Our results are generally consistent with their findings; however, in our XRD pattern, the t-phase is already predominant accompanied by a low amount of m-phase at ambient pressure shown in Fig.2(a) Furthermore, we observe that the monoclinic-to-tetragonal phase transition begins as early as $\sim$1 GPa and is completed at $\approx$ 10 GPa, indicating a lower pressure threshold for the complete transformation compared to previously reported results for 3YSZ \cite{Alzyab1987}.

With further pressure increase up to 30 GPa no major changes ($i.e.$ appearance or disappearance of Bragg peaks) that will signify a possible major pressure-induced phase transition are observed. However, a more detailed examination of the patterns reveals a progressive merging of certain Bragg peaks, with increasing pressure, see Fig.3(b). This merging of the relevant reflection doublets, resulting to  only one Bragg peak above $\approx$  18 GPa, signals the transformation towards the t''-or the c-phases. Following the methodology described by Lamas \textit{et al}.\cite{Lamas2000}, the gradual merging of reflections [220] and [004] indicates a phase transition from t-phase towards t''/c-phase. This transition is manifested by the single Bragg peak with merged [220]$_{t''}$ and [004]$_{t''}$ or a single [400]$_f$ reflection. As illustrated in Fig.3 (b), the merging begins at approximately 12 GPa and concludes around 18 GPa, signifying the t-to-t''/c-phase transformation. In other words, a pressure-induced increase of the symmetry towards a full cubic (a=b=c) or a pseudocubic tetragonal (a=b=c/$\sqrt{2}$) t''-phase. As mentioned before, the only difference between t''- and c- phases is a slight displacement of O atoms in the case of the t''-phase, that is beyond the probing ability of our powder XRD measurements \cite{Stavrou2012}.  

\begin{figure}[h]
    \centering
    \includegraphics[width=\linewidth]{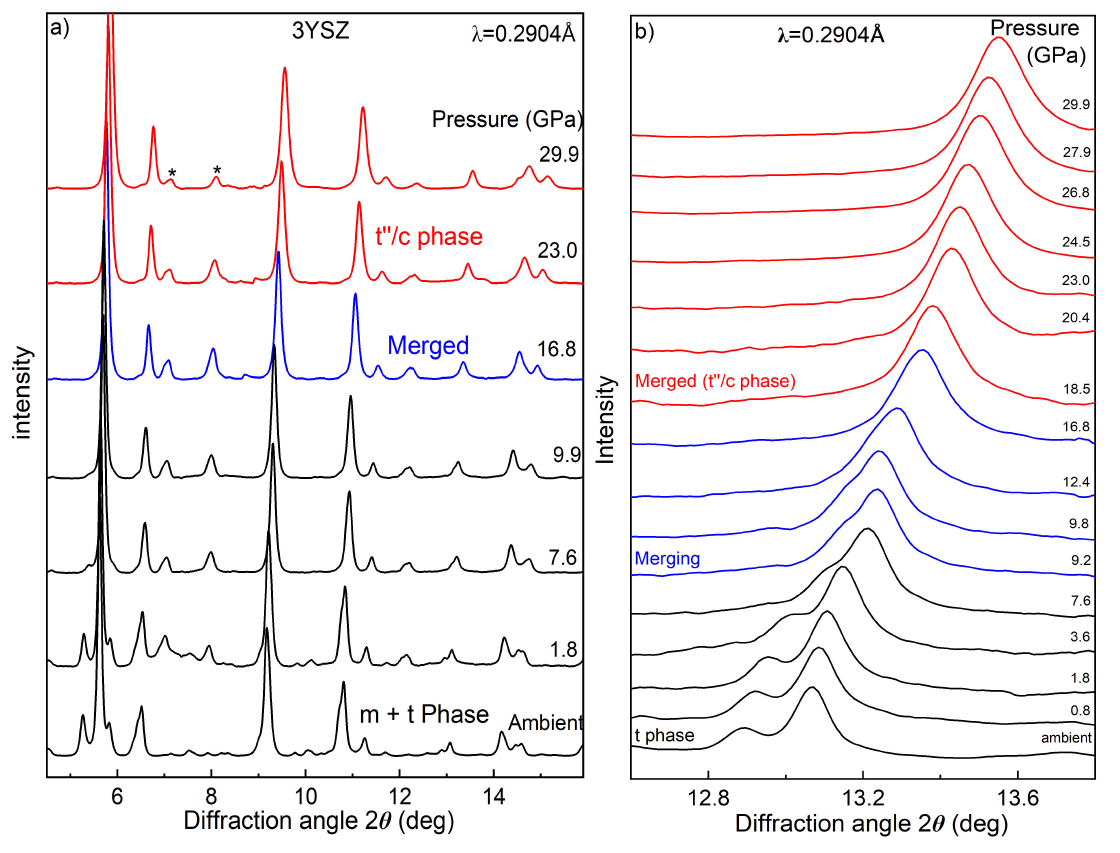}
    \caption{(a) XRD patterns for 3YSZ up to 30 GPa. (b) XRD patterns in the diffraction angle 2$\theta$ region of the  [220]$_t$, [004]$_t$, [400]$_f$ reflections for 3YSZ up to 30 GPa.}
\end{figure}

\subsubsection{XRD measurements of 8YSZ up to 40 GPa}
Selected XRD patterns of 8YSZ at various pressures up to 40 GPa are shown in Fig.4. First, we note the gradual disappearance of the m-phase that is concluded ($i.e.$ disappeared) at $\approx$ 10 GPa, in agreement with the findings in the case of 3YSZ. A closer inspection of the patterns at very low pressure (even at initial compression of 0.2 GPa), see Fig. 4(b), reveals the appearance of low intensity Bragg peaks that cannot be explained by the t''-phase but only by the t-phase (that was absent at ambient conditions). Thus, it is plausible to assume that the m-phase first transforms to the t-phase (as also observed for 3YSZ) resulting to a practically coexistence of three phases (m, t and t'') up to 10 GPa. Above this pressure only a mixture of t- and t''-phases remains. With further pressure increase above 15 GPa, a merging of the characteristic Bragg peaks (see discussion in the case of 3YSZ) is observed, that is concluded at $\approx$ 35 GPa. 

\begin{figure}[h]
    \centering
    \includegraphics[width=\linewidth]{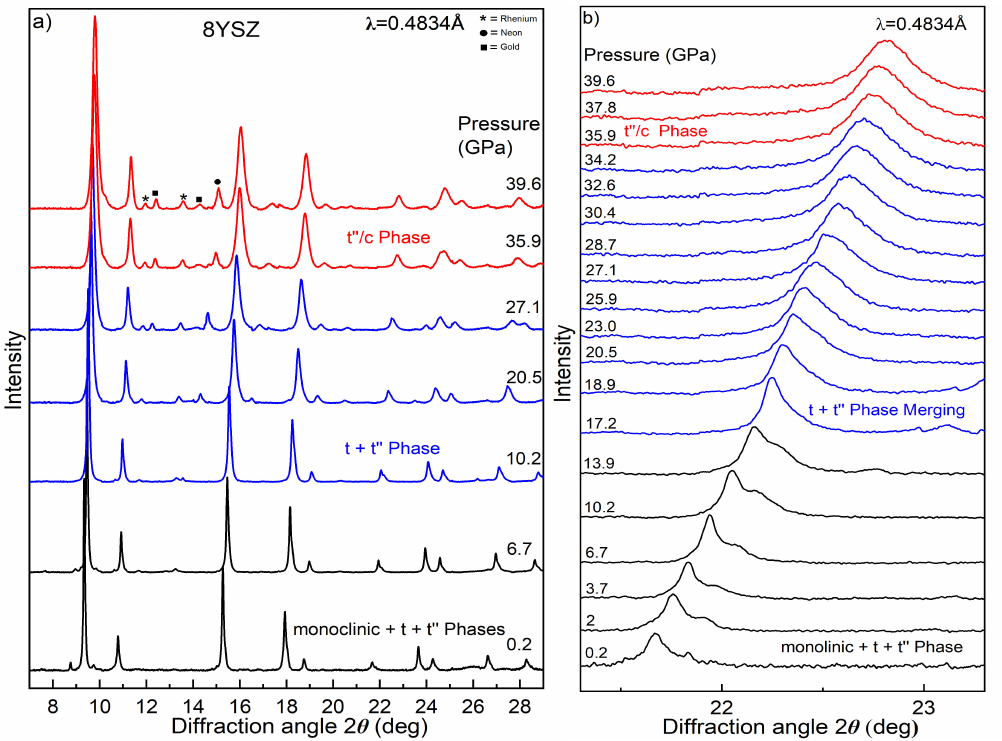}
    \caption{(a) XRD patterns for 8YSZ up to 40 GPa. The peaks marked with asterisks, squares and cycles originate from  Re (gasket material), Au (pressure marker) and Ne (PTM), respectively (b) XRD patterns in the diffraction angle 2$\theta$ region of the [220]$_t$, [004]$_t$, [400]$_f$ reflections for 8YSZ up to 40 GPa.} 
\end{figure}

Above 35 GPa, the XRD patterns are in agreement with  the existence of only the t''- or c-phases. Although this is in agreement with the observation in the case of 3YSZ, we note that  the fact that  the m-phase in 8YSZ initially transforms into the t-phase results to : a) coexistence of 3 phases at lower pressures and b)   shifts the complete t $\rightarrow$ t''/c transformation towards higher pressure. 

\subsection{RS measurements}
As discussed at the beginning of the results section, the large difference in X-ray scattering cross-sections between the heavy elements Zr (Z = 40) and Y (Z = 39) compared to the much lighter O (Z = 8) renders the displacement of oxygen atoms within the unit cell difficult to detect using only XRD. This limitation is particularly critical when attempting to distinguish between the  t''-and the c-phases, which primarily differ in local oxygen ordering (oxygen displacements). Therefore, Raman spectroscopy, which is highly sensitive to local order modifications, is required for further investigation. According to group theory, the Raman active zone center modes  are $\widetilde{\Gamma}_{R}$ = $A_{1g}$+2$B_{1g}$+3$E_g$ for the t/t''-phases  and $\widetilde{\Gamma}_{R}$ = $T_{2g}$  for the c-phase \cite{Kroumova2003}. We reiterate that Raman spectroscopy cannot distinguish between t and t''phase, given that group theory predicts same numbed and symmetries of active Raman modes. 

\subsubsection{RS measurements for 3YSZ}
Selected Raman spectra  of 3YSZ at various pressures up to 40 GPa are shown in Fig.5. In the case of 3YSZ, Raman spectroscopy confirms the coexistence of m -and t-phases at low pressures and the gradual m $\rightarrow$ t transition that is concluded up to 10 GPa, as documented by the gradual disappearance of the characteristic Raman modes of the m-phase (asterisks in Fig. 5(a)).  As the pressure increases, the two peaks at $\approx$ 150cm$^{-1}$ and 250cm$^{-1}$   gradually merge. This behavior matches the expected features of the t''-phase \cite{Hemberger2016}, and corresponds well to our  XRD results.  At higher pressures, adopting the analysis by Yashima \textit{et al}. \cite{Yashima1993b}, we probe the intensity ratio I4/I6, that corresponds to   the intensities of the 4th and 6th peaks of the Raman spectrum of the t/t''-phases, see downward arrows n Fig, 5(a). According Ref. \cite{Yashima1993b} the intensity ratio of the 4th and 6th peaks provides a reliable indicator of the tetragonal-cubic phase transition/boundary, underlying the fact that the only Raman active mode of the c-phase ($T_{2g}$) will be the continuation of the 6th peak of the t/t''-phases. 

\begin{figure}[h]
    \centering
    \includegraphics[width=\linewidth]{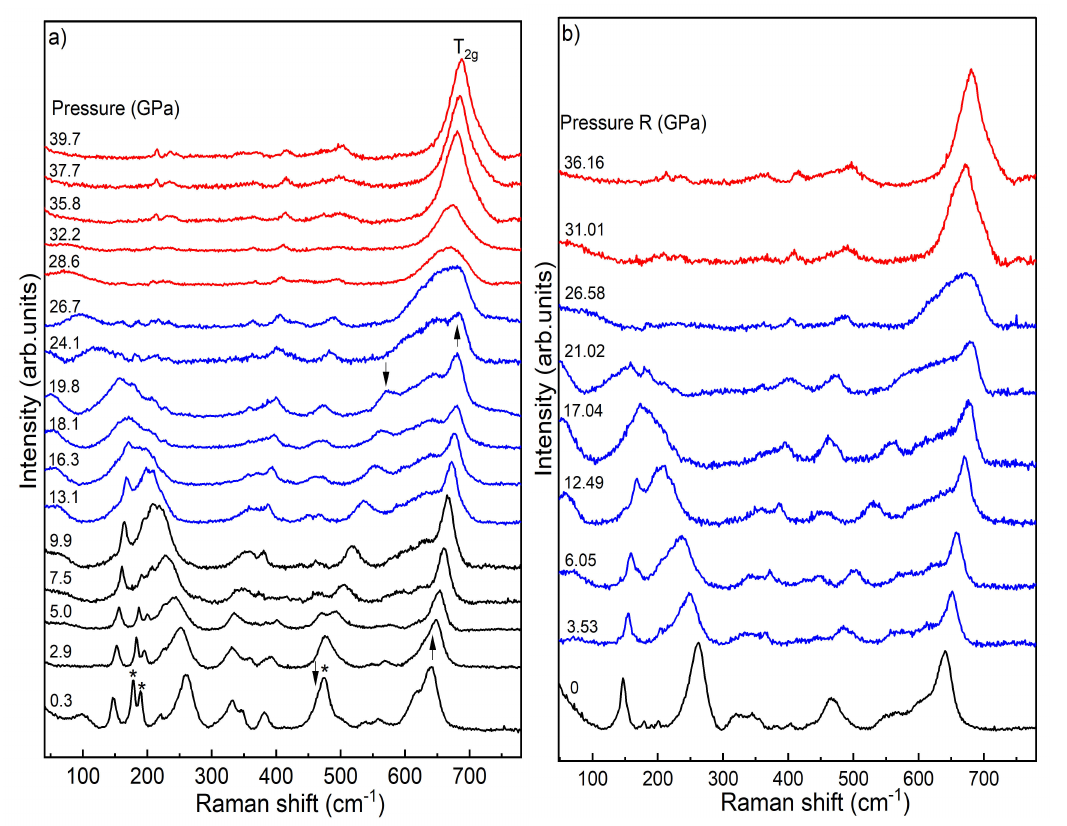}
    \caption{Raman spectra of 3YSZ upon (a) pressure increase from ambient to 40 GPa and (b) pressure decrease up to ambient pressure. In (a) the asterisks indicate the strongest Raman modes of the monoclinic phase \cite{Hemberger2015,Loganathan2012,torres2009}. The   downward and upward arrows correspond to the 4th and 6th peaks of the Raman spectrum of the t-phase, respectively, following Yashima \textit{et al.}\cite{Yashima1993b} analysis.  }
\end{figure}

In our study, the I4/I6 ratio decreases progressively with pressure and reaches a minimum at $\approx$ 26 GPa. This observation is similar to the previous analysis  by Yashima $et$ {} $al.$, and can be attributed to the tetragonal-to-cubic phase transition.  Above $\approx$ 28 GPa, the reduction of relatively low-intensity modes and the appearance of a dominant high-intensity peak reveal the transition towards the fluorite-typy structure, characterized by a single T$_{2g}$ active mode in agreement with previous studies on cubic YSZ \cite{Yashima1994, bouvier2000, ishigame1987, torres2009, Hemberger2015, Hemberger2016}, and hence mark the transition towards the fluorite-cubic structure. Thus, Raman spectroscopy results clarify that the high-pressure ($i.e.$ above 28 GPa)  phase of 3YSZ is the c-phase, instead of the t''-phase. 

Upon decompression (Fig.5 (b)), the c-phase remains stable down to $\approx$ 26 GPa. Below 26 GPa the c-phase transforms to t-phase. The t-phase remains dominant down to ambient conditions with minor presence of m-phase which will be discussed in details in the later section. Overall, the Raman spectra upon decompression  document the reversibility of the observed phase transitions, with minimum hysteresis.  This consistency suggests a second-order and reversible transition from the t-phase to the c-phase.

\subsubsection{RS measurements for 8YSZ}
Selected Raman spectra  of 8YSZ at various pressures up to 39 GPa are shown in Fig.6. Similar to 3YSZ, Raman spectroscopy confirms the coexistence of m- and t/t''-phases at low pressures and the gradual m $\rightarrow$ t/t''. As the pressure increases the gradual m → t/t” transition is observed below ~12 GPa. Further compression results in decrease in ratio I4/I6. This observation is similar to the previous analysis and by Yashima $et${} $al.$ , and can be attributed to the tetragonal-to-cubic phase transition.   Above 31 GPa,   the remaining  of a single Raman peak indicates the phase transitions towards the c-phase fluorite-cubic structure. However, in contrast to the  3YSZ case, instead of a dominant and relatively sharp high-intensity peak corresponding to the  T$_{2g}$ active mode at around 650cm$^{-1}$, a weaker broader peak is observed. This can be attributed to a defected structure, in comparison to the  perfect fluorite structure, owing to the creation of oxygen vacancies upon higher doping with Y$_2$O$_3$ \cite{kim1997}.  Furthermore, the weaker  and broader Raman modes of the t-phase may also result from the higher concentration of oxygen vacancies in the 8YSZ sample. 

\begin{figure}[h]
    \centering
   \includegraphics[width=\linewidth]{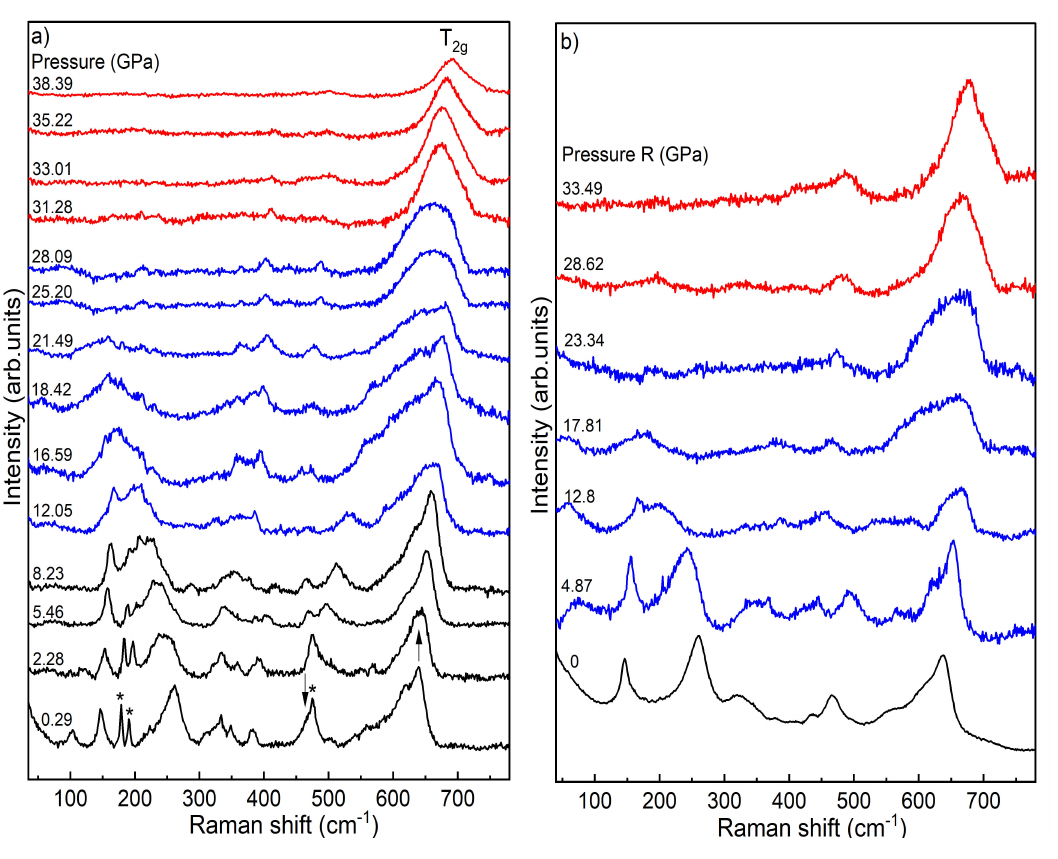}
    \caption{Raman spectra of 8YSZ upon (a) pressure increase from ambient to 38 GPa and (b) pressure decrease up to ambient pressure. In (a) the asterisks indicate the strongest Raman modes of the monoclinic phase \cite{Hemberger2015,Loganathan2012,torres2009}. The   downward and upward arrows correspond to the 4th and 6th peaks of the Raman spectrum of the t-phase, respectively, following Yashima \textit{et al.}\cite{Yashima1993b} analysis.}
\end{figure}

Upon decompression, the  Raman spectra in Fig.6(b)  show similar evolution to the 3YSZ case, $i.e.$ all phase transitions are found to be reversible with negligible hysteresis.

\subsubsection{XRD and RS measurements after full pressure release. }
After full pressure release to ambient conditions, both 3YSZ and 8YSZ exhibit full reversibility  to the initial t- and t''-phase, respectively, as evident from  both XRD and RS measurements, see Fig. 7(a),(b). However, it shall be noted that the intensities of the characteristic Bragg peaks and Raman modes of the m-phase for both 3YSZ and 8YSZ become exceedingly weak, indicating that the amount of recovered m-phase is also minuscule, if any amount of this phase exists at all.

\begin{figure}[h]
    \centering
   \includegraphics[width=\linewidth] {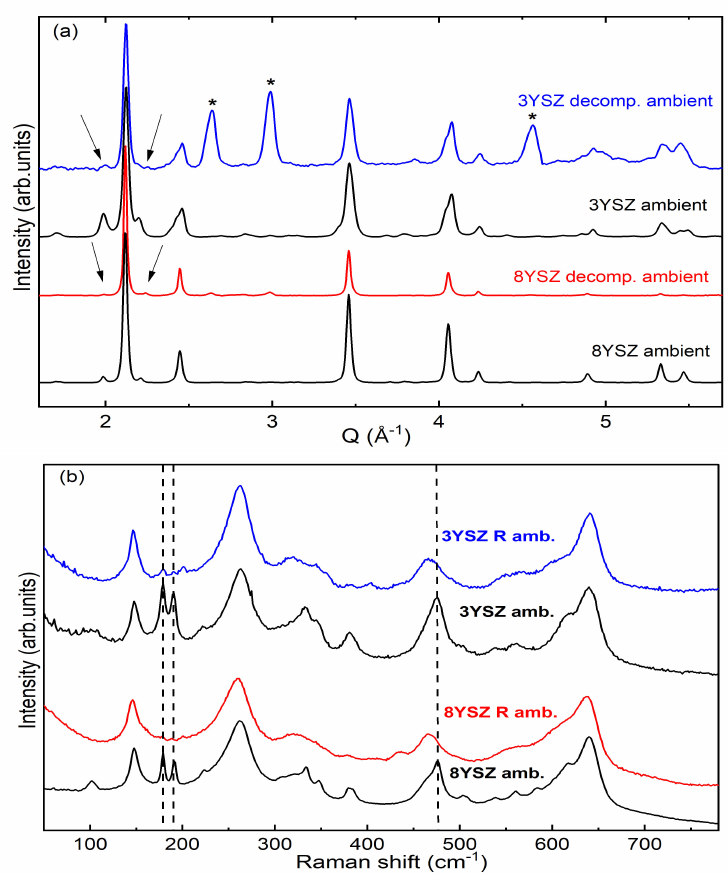}
    \caption{Comparison between (a) XRD patterns and (b) Raman spectra at ambient conditions and after pressure release for 3YSZ and 8YSZ. The arrows in (a) and the vertical dashed lines in (b) denote the Bragg peaks and the Raman modes of the m-phase, respectively. The asterisks in (a) denote  Bragg peaks originating from Re (gasket material).}
\end{figure}

\section{Discussion}

By combining the results of both XRD and Raman measurements, the high-pressure structural evolution of the 3YSZ and 8YSZ can be summarized in Fig. 8. The main aspects are as follows: a) For 3YSZ a coexistence of the t-(predominant) and the m-phases is observed at ambient conditions. Upon initial pressure increase the m-phase gradually transforms towards the t-phase, a transition that is concluded  up to 10 GPa.  With further pressure increase, the t-phase gradually transforms to the t''-phase, a transformation that is concluded above 18 GPa. At higher pressures,  the t''-phase  transforms to the c-phase above 28 GPa, that remains stable up to, at least, 40 GPa.

\begin{figure}[h]
    \centering
   \includegraphics[width=\linewidth]  {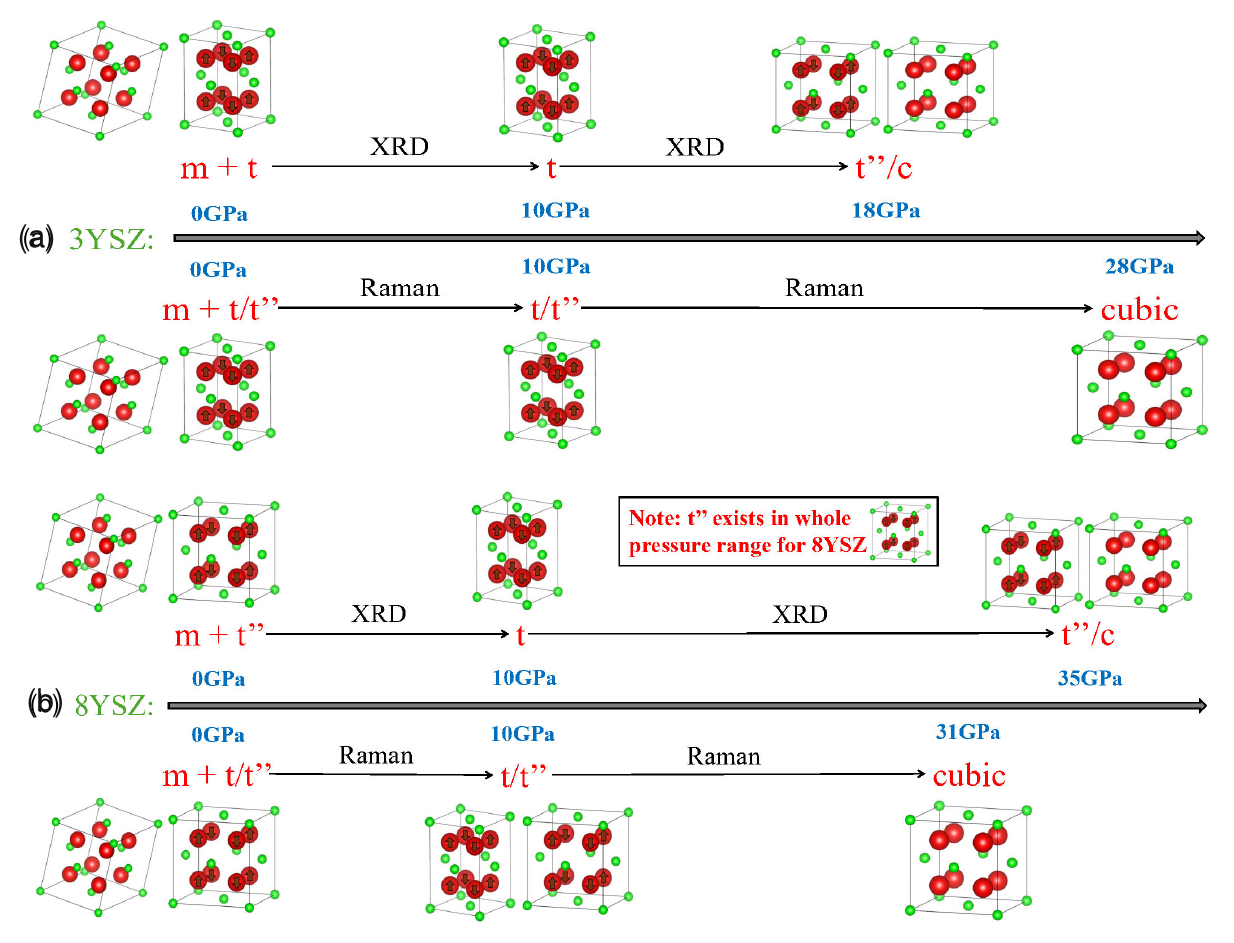}
    \caption{High-pressure structural evolution of (a) 3YSZ and (b) 8YSZ as determined by the XRD and RS measurements of this study, see text for details.}
\end{figure}

b) For 8YSZ a coexistence of the t''-(predominant) and the m-phases is observed at ambient conditions. Upon initial pressure increase the m-phase gradually transforms towards the t-phase, resulting to a coexistence of 3 phases.  At 10 GPa,  the m is  fully transformed to the t-phase. The t- and t''-phases continue to coexist up to 31 GPa. Above this pressure, both transform towards the c-phase, that remains stable up to, at least, 39 GPa. 

For both the 3YSZ and 8YSZ, pressure increase promotes phase transitions  towards higher symmetry structures: monoclinic to tetragonal and tetragonal to cubic. Moreover, the increase of Y$_2$O$_3$ doping appears to increase the critical pressure for the transition towards the ideal fluorite structure, although the difference is relatively small to draw solid conclusions. Upon pressure decrease, all above mentioned phase transitions are fully reversible with negligible hysteresis, with the exception of the practical disappearance of the m-phase at ambient conditions. 

Having clarified the exact crystal structures at ambient conditions, the XRD patterns were indexed and the corresponding lattice parameters and cell volume  are provided   in Table I, showing an agreement  with relevant previous studies. Likewise, the lattice parameters and cell volume as a function of pressure were determined and are plotted in Fig. 9 and summarized in Table I.

\begin{figure}[h]
    \centering
    \includegraphics[width=\linewidth] {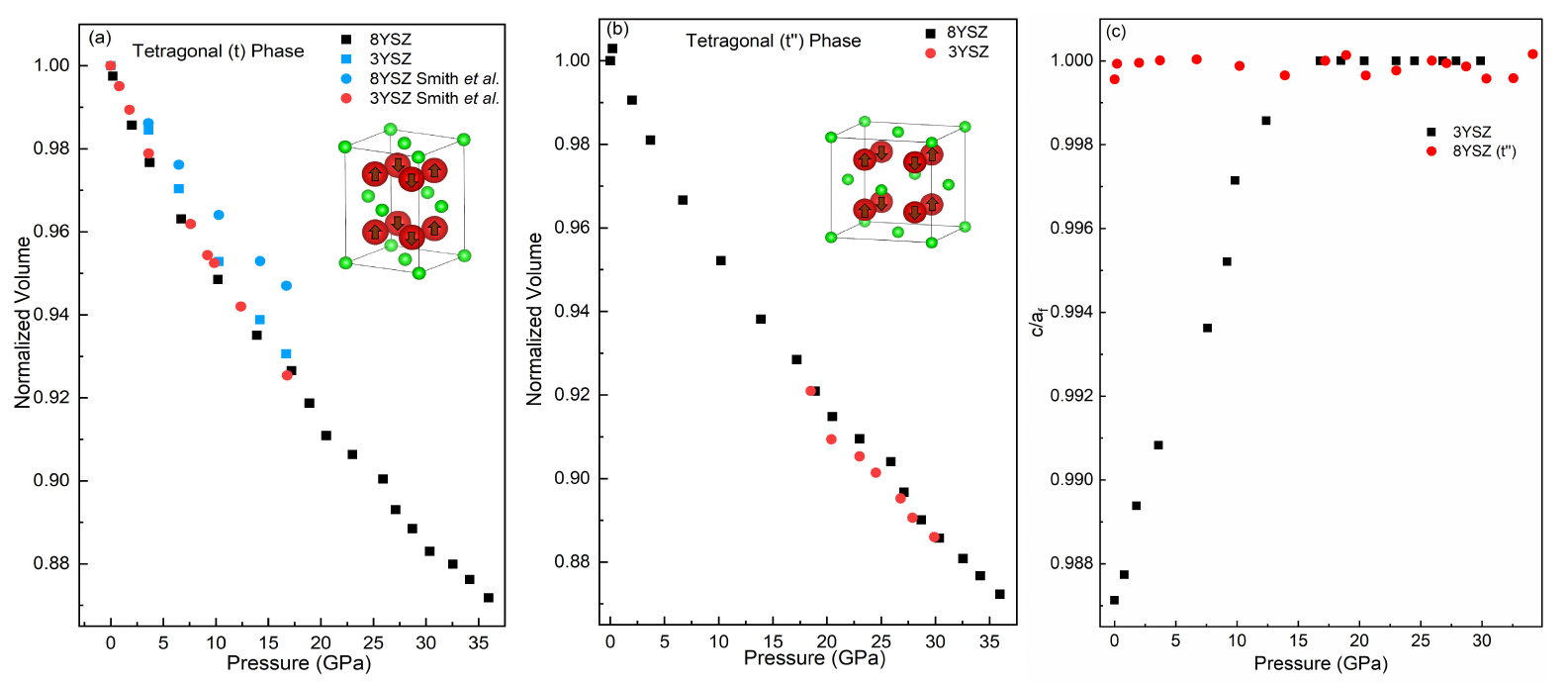}
    \caption{Pressure dependence of the   unit cell volume of (a) the tetragonal (t) phase and  (b) the  tetragonal (t'') phase of 3YSZ and 8YSZ. The P-V data  from Ref. \cite{smith2016} are also provided in (a) for comparison. (c) c/a$_f$ versus pressure of  3YSZ and 8YSZ.  }
\end{figure}

\begin{table}[h]
\centering
\scriptsize
 \caption{Experimental structural parameters of 3YSZ and 8YSZ  at selected pressures: phase, lattice parameters, unit cell volume,  Bulk modulus \textit{B} and its pressure derivative \textit{B'} (as determined by fitting a 3$^{rd}$ order Birch-Murnaghan  EOS \cite{Birch1978}  to the  experimental data) at the  onset pressure. The structural parameters from and previous studies at ambient conditions are also provided.}
\begin{tabular}{ccccccc}
\hline
              &                &                        &          \textbf{3YSZ}              &                            &         &       \\ \hline
Pressure (GPa)&Phase    & a (\AA) & c (\AA) & V (\AA$^3$) & B (GPa) & B'    \\ \hline
0             & \textit{t}            & 3.611                 & 5.173                   & 67.42                    & 170(4)  &6.4(7)     \\ 
   0      &          & 3.602                  & 5.179                    & 67.19                      &\cite{Wang1999} \\ 
3.6           &                       &3.588	              &5.126	                &65.99                     &         &       \\ 
7.6           &                       &3.571	              &5.085	                &64.84	                   &         &       \\ 
12.4          &                       &3.549	              &5.042	                &63.51                   &         &       \\ \hline
18.5          & \textit{t''}          &3.528	&4.989	&62.01	&&\\ 
24.5          &                       &3.503	&4.953	&60.79	                    & 258(18)        &  4(fixed)     \\ 
26.8          &                  &3.495	&4.942	&60.35	                      &         &       \\ \hline
27.9          &         \textit{c}       &               4.934	&4.934&	120.08	                   & 	&   \\ 
29.9          &                      &4.925	&4.925	&119.46                    &         &       \\ \hline \hline			
    &                &                        &          \textbf{8YSZ}              &                            &         &       \\ \hline
0&	\textit{t''}  	&3.633&	5.136&	67.8&	189(4)&	4.8(4)\\
0& &3.635                  & 5.141                    & 67.93                      & \cite{Yashima1994} \\
6.7&		&3.594&	5.079	&65.54	&	& \\
13.9	&	&3.557&	5.028&	63.61	&	& \\
20.5	&&	3.527	&4.986	&62.03& & \\		
28.7	&&	3.495	&4.941	&60.35	&&\\		\hline
6.7	 &\textit{t} &	3.594	&5.055&	65.3	& 178(8)&	7.8 (8)\\
13.9	&&	3.558	&5.007	&63.4	&	&\\
20.5	&&	3.528	&4.961	&61.76	&	&\\
28.7	&&	3.495	&4.932	&60.24	&	&\\\hline
35.9	&\textit{c} 	&4.907	&4.907	&118.43	&&\\
39.6	&	&4.896	&4.896	&117.37	&&	\\\hline
\end{tabular}
\end{table}

The P-V data  for both specimens exhibit smooth and continuous transitions from the t- to t''-phase, indicating a second-order phase transition, as shown in Fig.9 (a) and (b). We have fitted  3$^{rd}$-order (for ambient or near ambient phases) or 2$^{nd}$-order (for high-pressure phases) Birch-Murnaghan EoSs \cite{Birch1978}  to the  experimental data, and the determined bulk moduli for the ambient tetragonal phases are listed in Table I. Although 3YSZ and 8YSZ adopt different tetragonal phases at ambient conditions, the results indicate that 8YSZ is slightly less compressible that 3YSZ, although it is expected to possess higher number of O vacancies. Comparison of the P-V data of this study with a previous EOS study of 3 and 8YSZ \cite{smith2016} up to 15 GPa, see Fig. 9(a), reveals a good overall agreement for 8YSZ. However, a deviation for 3YSZ is observed, that can be attributed to the   substantial non-hydrostatic conditions (no PTM) in Ref. \cite{smith2016}. 

Fig.9 (c) illustrates the pressure dependence of the c/a$_f$ ratio for the stable at ambient conditions t-phase in 3YSZ and the t''-phase in 8YSZ, where a$_f$ = $\sqrt{2}$a. For 8YSZ, as expected, the c/a$_f$ for the pseudo-cubic t''-phase slightly fluctuates around 1. This slight fluctuation is due  to the fact that the identification of exact position of the Bragg peak   for 8YSZ  is challenging, due to the overlap between the reflections of the minor t-and the dominant t''-phases, leading to uncertainty in quantifying the lattice parameters of the t''-phase. On the other hand, the c/a$_f$ of the t phases in 3YSZ shows a steep increase towards the ideal c/a$_f$=1 upon initial compression, indicating a progressive but rapid symmetrization.  Above 18 GPa, given the already discussed t $\rightarrow$ t'' transition,  c/a$_f$ reaches and retains the ideal c/a$_f$=1 value. 

Finally, we note that pressure and temperature usually have opposite effect on the symmetry of observed crystal structures. Our study documents that increase of temperature in pure ZrO$_2$ and of pressure in YSZ both can  stabilize the c-phase. The role oxygen vacancies in this observation shall be further examined.

\section{Conclusion}
The structural evolution of 3YSZ and 8YSZ under pressure has been studied using XRD and RD measurements up to 40 GPa at RT using Ne as PTM. The combination of the results from both techniques allowed an accurate determination of the structural evolution under pressure. For both specimens, pressure promotes higher symmetry phases: a) the minority at ambient conditions monoclinic phase transforms towards the tetragonal t-phase upon initial compression and b) t/t''-phases transform  towards the c-phase upon further compression. The c-phase remains stable up to the highest pressure of this study (40 GPa). Upon pressure release all pressure-induced phase traditions  are fully reversible with negligible hysteresis, underscoring the second-order nature of the phase traditions, with the exception of the practical  disappearance  of the m-phase in the fully decompressed specimens. This study underscores the need for a combined XRD and RS study in the case of closely related and competing crystal structures under pressure. 

\begin{acknowledgments}
The work performed at GTIIT was supported by funding from the Guangdong Technion Israel Institute of Technology and the Guangdong Provincial Key Laboratory of Materials and Technologies for Energy Conversion, MATEC (Grant No. 2022B1212010007, Guangdong Department of Science and Technology). We thank the financial support from the Technion-GTIIT seed grant program. Part of the experiments was carried out at BL10XU of SPring-8 with the approval of the Japan Synchrotron Radiation Research Institute (JASRI) (Proposals No.2024A1096 and 2024B1182). Additional synchrotron radiation experiments were performed at PETRA III, beamline P02.2, with the support of DESY (Hamburg, Germany), a member of the Helmholtz Association HGF. Beamline 12.2.2 at the Advanced Light Source is a DOE Office of Science User Facility under contract no. DE-AC02-05CH11231. We thank the Shanghai Synchrotron Radiation Facility of BL17UM(31124.02.SSRF.BL17UM) and BL15U1 (31124.02.SSRF.BL15U1) for the assistance on high-pressure XRD measurements.
\end{acknowledgments}

\end{document}